\documentclass[aps,pra,twocolumn,amsfonts,amsmath,showpacs,floatfix,footinbib,groupedaddress,superscriptaddress]{revtex4}
\usepackage{amssymb,graphics,graphicx,times,bm,bbm,multirow}

\begin{document}

\title{Efficient estimation of nearly sparse many-body quantum Hamiltonians}
\author{A. Shabani}
\affiliation{Department of Chemistry, Princeton University,
Princeton, New Jersey 08544}
\author{M. Mohseni}
\affiliation{Research Laboratory of Electronics, Massachusetts
Institute of Technology, Cambridge, MA 02139}
\author{S. Lloyd}
\affiliation{Research Laboratory of Electronics, Massachusetts
Institute of Technology, Cambridge, MA 02139}
\affiliation{Department of Mechanical Engineering, Massachusetts
Institute of Technology, Cambridge, MA 02139}
\author{R. L. Kosut}
\affiliation{SC Solutions, Sunnyvale, CA 94085}
\author{H. Rabitz}
\affiliation{Department of Chemistry, Princeton University,
Princeton, New Jersey 08544}

\begin{abstract}
We develop an efficient and robust approach to Hamiltonian
identification for multipartite quantum systems based on the method
of compressed sensing. This work demonstrates that with only
$\mathcal{O}(s \log(d))$ experimental configurations, consisting of
random local preparations and measurements, one can estimate the
Hamiltonian of a $d$-dimensional system, provided that the
Hamiltonian is nearly $s$-sparse in a known basis.  We numerically
simulate the performance of this algorithm for three- and four-body
interactions in spin-coupled quantum dots and atoms in optical
lattices.  %%%masoud change
Furthermore, we apply the algorithm to characterize Hamiltonian fine
structure and unknown system-bath interactions. %%% end change

\end{abstract}
\maketitle
\section{introduction}
The dynamical behavior of multipartite quantum systems is governed by
the interactions amongst the constituent particles. Although, the
physical or engineering considerations may specify some generic properties about
the nature of quantum dynamics, the specific form
and the strength of multi-particle interactions are typically
unknown. Additionally, quantum systems usually have an unspecified
interaction with their surrounding environment. In principle, one
can characterize quantum dynamical systems via
\textquotedblleft quantum process tomography" (QPT) 
\cite{dcqd3,dcqd31,dcqd32,dcqd33,dcqd34,dcqd35,dcqd36,dcqd37}.
However, the relationship between relevant physical properties of a
system to the information gathered via QPT is typically unknown.
 Alternatively, knowledge about the
nature of inter- and intra- many-body interactions within the system
and/or its environment can be constructed by identifying a set of (physical
or effective) Hamiltonian parameters generating the dynamics
\cite{JM,Cory,Cory2,Cory3,Cole,Cole2,Soph,Levi,Mohseni09,Franco}. Currently,
a scalable approach for efficient estimation of a full set of Hamiltonian
parameters does not exist.

The dynamics of a quantum system can be estimated by observing the
evolution of some suitable test states. This can be achieved by a
complete set of experimental configurations consisting of
appropriate input states and observables measured at given time
intervals. Knowledge about the dynamics may then be reconstructed
via inversion of the laboratory data by fitting a set of dynamical
variables to the desired accuracy. Estimating Hamiltonian parameters
from such a procedure faces three major problems: (1) The number of
required physical resources %experimental configurations
grows exponentially with the degrees of freedom of the system
\cite{dcqd3,dcqd31,dcqd32,dcqd33,dcqd34,dcqd35,dcqd36,dcqd37}. (2) There are
inevitable statistical errors associated with the inversion of
experimental data \cite{dcqd3,dcqd31,dcqd32,dcqd33,dcqd34,dcqd35,dcqd36,dcqd37}.
(3) The inversion generally involves
solving a set of nonlinear and non-convex equations, since the
propagator is a nonlinear function of Hamiltonian parameters
\cite{JM,Cory,Cory2,Cory3,Cole,Cole2,Soph,Levi,Mohseni09,Franco}.
The first two problems are always present with any
form of quantum tomography, but the last problem is specific to the
task of Hamiltonian identification as we wish to reconstruct the
generators of the dynamics. Many quantum systems involve two-body
local interactions, so the goal is often to estimate sparse
Hamiltonians with effectively a polynomial number of unknown
parameters. Unfortunately, quantum state and process tomography
cannot readily exploit this potentially useful feature.

%%%masoud change
The highly nonlinear feature in the required inversion of laboratory
data was studied in Ref.\cite{JM} in which closed-loop learning
control strategies were used for the Hamiltonian identification. In
that approach one estimates the unknown Hamiltonian parameters by
tailoring shaped laser pulses to enhance the quality of the
inversion. Identification of time-independent (or piece-wise
constant) Hamiltonians have been studied for single-qubit and
two-qubit cases \cite{Cole,Cole2} to verify the performance of quantum
gates. Estimation of these Hamiltonians is typically achieved via
monitoring the expectation values of some observable, e.g.
concurrence, which are time periodic functions. Through Fourier
transform of this signal the identification task is reduced to
finding the relative location of the peaks and heights of the
Fourier spectrum \cite{Cole,Cole2}. Bayesian analysis is another method
proposed for robust estimation of a two-qubit Hamiltonian \cite{Soph}.
The difficulty with these methods is
then scalability with the size of the system. A symmetrization
method for efficient estimation of the magnitude of effective
two-body error generators in a quantum computer was studied in
\cite{Levi} by monitoring quantum gate average fidelity decay.
Recently, it was demonstrated that direct or selective QPT schemes
could be used for efficient identification of short-time behavior of
sparse Hamiltonians \cite{Mohseni09} assuming controllable two-body
quantum correlations with auxiliary systems and the exact knowledge
of the sparsity pattern. Another scheme for the determination of the
coupling parameters in a chain of interacting spins with restricted
controllability was introduced in Ref. \cite{Franco}.

%%%end change

In this work, inspired by recent advances in classical signal
processing known as \emph{compressed sensing} \cite{CS.intro}, we
use random \textit{local} input states and measurement observables
for efficient Hamiltonian identification. We show how the
difficulties with the nonlinearity of the equations can be avoided
by either a short time or a perturbative treatment of the dynamics.
We demonstrate that randomization of the measurement observables
enables compressing the extracted Hamiltonian information into a
exponentially smaller set of outcomes. This is accomplished by a
generalization of compressed sensing to utilize random matrices with
correlated elements. This approach is applicable for Hamiltonians
that are nearly sparse in a known basis with an arbitrary unknown
sparsity pattern of parameters. The laboratory data can then be
inverted by solving a convex optimization problem. This algorithm is
highly tolerant to noise and experimental imperfections. The power
of this procedure is illustrated by simulating three- and four-body
Hamiltonians for neutral atoms in an optical lattice and
spin-coupled quantum dot systems, respectively.
%%%masoud change
Furthermore, we directly apply the algorithm to estimate Hamiltonian
fine structure and characterize unknown system-bath interactions for
open quantum systems. %%%end change

\section{Quantum dynamical equations}The time evolution of a quantum
system in a pure state is governed by the Shr\"{o}dinger equation, $%
d\left|\psi(t)\right\rangle/dt =-iH\left|\psi(t)\right\rangle $. The
solution of this equation for a time-independent Hamiltonian can be simply
expressed as $\left|\psi(t)\right\rangle
=\exp(-itH)\left|\psi(0)\right\rangle $. In principle, the Hamiltonian of the
system $H$ can be estimated by preparing an appropriate set of test states
$\{\left|\psi_k\right\rangle\}$ and measuring the expectation value of a
set of observables $\{M_j\}$ after the system has evolved for a certain
period of time. The expectation value of these observables can be expressed as
\begin{equation}
p_{jk}=\langle M_j\rangle_{\psi_k}=\left\langle\psi_k\right| e^{itH}M_j
e^{-itH}\left|\psi_k\right\rangle  \label{nonlinear}
\end{equation}

Equation (\ref{nonlinear}) implies that the experimental outcomes $%
\{p_{jk}\}$ are nonlinear functions of the Hamiltonian parameters. To
avoid the difficulties of solving a set of coupled nonlinear equations we
consider the short time behavior of the system. Monitoring the short time
dynamics of the system is valid when the relevant time scales of the system
evolution satisfy $t \ll K^{-1}$ where, for positive operator-valued measure (POVM) operators $\{M_j\}$,
the constant $K$ equals $2||H||_{spec}$.
The general expression of $K$ is given in appendix B,
also see appendix A for definition of the norms.
This yields the linearized form of the Eq. (\ref{nonlinear})
\begin{eqnarray}
p_{jk}=\left\langle \psi_{k}\right|M_{j}\left|\psi_{k}\right\rangle
+it\left\langle \psi_{k}\right|[H,M_{j}]\left|\psi_{k}\right\rangle+\mathcal{O}(K^2t^2)
\label{linear}
\end{eqnarray}

The linear approximation contains enough information to
fully identify the Hamiltonian and the higher order terms do not
provide additional information. The short-time approximation implies
prior knowledge about the system dynamical time-scale or the order of
magnitude of $||H||_{spec}$. This prior knowledge can be available
from generic physical and engineering considerations.
For example, in solid-state quantum
devices the time-scale of single qubit rotations is typically on the
order of 1-10 ns. The switching time for exchange interactions
varies among different solid-state systems from 1ps to 100ps, (for
more details see appendix B.)

We expand the Hamiltonian in an orthonormal basis $
\{\Gamma_{\alpha}\}$, where $\text{Tr}(\Gamma^{\dagger}_{\alpha}\Gamma_{
\beta})=d\delta_{\alpha,\beta}$: $H=\sum_{\alpha} h_{\alpha}\Gamma_{\alpha}$
. Here $d$ is the dimension of the Hilbert space. In this representation the
Hamiltonian parameters are the coefficients $h_{\alpha}$. The expanded form
of the above affine equation (\ref{linear}) is
\begin{eqnarray}
\bar{p}_{jk}=it\sum_{\alpha}\left\langle
\psi_{k}\right|[\Gamma_{\alpha},M_{j}]\left|\psi_{k}\right\rangle h_{\alpha}
\label{expanded}
\end{eqnarray}
Here we introduce the experimental outcomes as $\bar{p}_{jk}=p_{jk}-\left\langle \psi_{k}\right|M_{j}\left|\psi_{k}\right\rangle$,
since $\left\langle \psi_{k}\right|M_{j}\left|\psi_{k}\right\rangle$ is
\textit{a priori} known. The relation (\ref{expanded}) corresponds to a single
experimental configuration ($M_j$,$\left|\psi_{k}\right\rangle$). For a $d$-dimensional system, the total number of
Hamiltonian parameters $h_{\alpha}$ is $d^2$. Thus, one requires the
same number of experimental outcomes, $p_{jk}$ that
leads to $d^2$ linearly independent equations.
For a system of $n$ qubits, this number
grows exponentially with $n$ as $d=2^{2n}$. In order to devise an efficient
measurement strategy we will focus on physically motivated \textit{nearly sparse}
Hamiltonians.

A Hamiltonian $H$ is considered to be $s$-sparse if it only contains
$s$ non-zero parameters $\{h_{\alpha}\}$. More generally, a
Hamiltonian $H$ is termed nearly $s$-sparse, for a threshold
$\eta$, if at most $s$ coefficients $h_{\alpha}$ ($H=\sum
h_{\alpha}\Gamma_{\alpha}$) have magnitude greater than $\eta h_{max}$
where $h_{max}=\max(h_{\alpha})$.
By definition, the sparsity is basis dependent. However, for local interactions, the basis in which the Hamiltonian
is sparse is typically known from physical or engineering considerations.

\section{Compressed Hamiltonian estimation} Our algorithm
is based on general methods of so-called compressed
sensing that recently have been developed in signal processing theory \cite%
{CS.intro}. Compressed sensing allows for condensing signals and
images into a significantly smaller amount of data,
and recovery of the signal becomes possible
from far fewer measurements than required by traditional methods.

Compressed sensing has two main steps: encoding and decoding. The
information contained in the signal is mapped into a set of laboratory data
with an exponentially smaller representation. This compression can be
achieved by randomization of data acquisition. The actual signal %M: needs to
can be recovered via an efficient algorithm %M: such as
based on convex optimization methods.
Compressed sensing has been
applied to certain quantum tomography tasks. Standard compressed
sensing has been directly used for efficient pseudothermal ghost
imaging \cite{GI:09,GI:091}.
Recently, a quadratic reduction in the total number of
measurements for quantum tomography of a low rank density matrix has
been demonstrated using a compressed sensing approach \cite{Gross:09}.

%In this work we utilize the mathematical formulation of compressed
%sensing %M:, similar to \cite{Shabani:09},
%for efficient estimation of
%physically motivated nearly sparse quantum Hamiltonians.
%As we mentioned above first we need to efficiently extract the signal information.
Here, we first describe how the Hamiltonian information is compressed into the
experimental data. The output of a single measurement is related to the
unknown signal (Hamiltonian parameters) through the relation (\ref{expanded}%
). Suppose we try $m$ different experimental configurations (i.e., $m$ different
pairs of $(M_j,\left| \psi_k\right\rangle)$). This yields a set of
linear equations
\begin{equation}
\overrightarrow{p^{\prime }}=\Phi\overrightarrow{h}  \label{linear-relation}
\end{equation}
where $\Phi$ is a $m\times d^2$ matrix with elements $\Phi_{jk,\alpha}=it/\sqrt{m}%
\left\langle
\psi_{k}\right|[\Gamma_{\alpha},M_{j}]\left|\psi_{k}\right\rangle$
(A
factor $1/\sqrt{m}$ is included for simplifying the proofs, appendix C). In
general $m$ has to be greater than or equal to $d^2$ in order to solve Eq. (%
\ref{linear-relation}). A Hamiltonian estimation attempt with $m<d^2$
seems impossible as we face an underdetermined system of linear equations
with an infinite number of solutions. However, any two $s$-sparse
Hamiltonians $h_1$ and $h_2$ still can be distinguished via a properly
designed experimental setting, if the measurement matrix $\Phi$ preserves the
distance between $h_1$ and $h_2$ to a good approximation:
\begin{equation}
(1-\delta_{s}) ||h_2-h_1||_{l_2}^{2} \leq ||\Phi (h_2-h_1)||_{l_2}^{2} \leq
(1+\delta_{s})||h_2-h_1||_{l_2}^{2}  \label{RIP}
\end{equation}
for a constant $\delta_s\in (0,1)$. A smaller $\delta_s$ ensures
higher distinguishability of $s$-sparse Hamiltonians. The inequality relation (\ref{RIP})
is termed a \textit{restricted isometry property} (RIP) of the matrix $\Phi$ \cite{RIP2}. We now
discuss how to construct a map $\Phi$ satisfying this inequality, and how
small the value of $m$ can be made.

%A common approach to establish RIP (\ref{RIP}) for a matrix $\Phi$
%is by introducing randomness in the elements of this matrix. This approach benefits from
%measure concentration properties of random matrices.
%In classical signal processing each element $\Phi_{jk,\alpha}$ can be
%independently selected from a random distribution such as Gaussian or
%Bernoulli. Whereas in the Hamiltonian estimation formulation
%($\ref{linear-relation}$) there is no freedom for independent selection
%of the $\Phi$ matrix elements.
The RIP (\ref{RIP}) for a matrix $\Phi$ can be established by employing the
measure concentration properties of random matrices.
In each experiment the test state
and the measurement observable can be drawn randomly from a set of
configurations $\{M_j,\left| \psi_k\right\rangle\}$ realizable in the laboratory.
The independent selection of $\left| \psi_k\right\rangle$ and $M_j$ leads to a matrix $\Phi$ with
independent rows but correlated elements $\Phi_{jk,\alpha}$ in each row.
Thus the standard results from compressed sensing theory are not applicable here (appendix C).

%In this work, we generalize the standard compressed sensing algorithm such that
%the necessity for independent randomness in all elements of the measurement matrix, $\phi$, can be avoided.
In contrast, here we derive a concentration inequality for a matrix with
independent rows and correlated columns
as the backbone for the RIP of our quantum problem in appendix C.
Using Hoeffding's inequality,
we show that for any Hamiltonian $h$ and
a random matrix $\Phi$ with column only correlations, the random variable $||\Phi h||^2$ is
concentrated around $||h||^2$ with a high probability, i.e. $%
\forall$ $0<\delta<1$
\begin{equation}
\text{Prob.}\{|||\Phi h||_{l_2}^2-||h||_{l_2}^2|\geq \delta ||h||_{l_2}^2\}\leq
2e^{-mc_0(\delta+c_1)^2}  \label{RIP2}
\end{equation}
for some constants $c_0$ and $c_1$.

Using the above inequality, now we can show how an
exponential reduction in the minimum number of the required
configurations can be achieved for Hamiltonian estimation. The inequality (\ref{RIP2}) is
defined for any $h$ while the inequality in the definition of
RIP, Eq.(\ref{RIP}), is for any $s$-sparse $h$. As
shown in Ref. \cite{RIP}, there is an inherent
connection between these two inequalities. It is proved that any
matrix $\Phi$ satisfying (\ref{RIP2}) has RIP with probability
greater than $1-2\exp(-mc_0(\delta_\frac{s}{2}+c_1)^2+s[\log(d^4/s)+\log(12e/\delta_\frac{s}{2})])$.
In addition, whenever $m\geq c_2s\log(d^4/s)$, for a sufficiently
large constant $c_2$ one can find a constant $c_3\geq 0$ such that
the likelihood of the RIP to be satisfied converges exponentially fast to unity as $1-2\exp(-c_3
m)$.

The set of experimental configurations defined by Eq (4), and the concentration properties given
by Eq (5) and (6) can be understood as encoding the information of
a sparse Hamiltonian into a space with a lower dimension.
Next we need to provide an efficient method
for decoding in order to recover the original Hamiltonian. The
decoder is simply the minimizer of the $l_1$ norm of the signal $h$%
. Implementing this decoder is a special convex optimization problem, which can be solved
via fast classical algorithms, yet not stricktly scalable. Furthermore, the encoding/decoding scheme is robust to
noisy data as $||p^{\prime }-\Phi h||_{l_2}\leq \epsilon$ where $\epsilon$
is the noise threshold. Note that $\epsilon$ includes the error of linearization (see Eq.(\ref{linear}))
that is $\mathcal{O}(\sqrt{m}Kt^2)$. Denote $h_0$ as the true representation of the
Hamiltonian. For a threshold $\eta$, $h_0(s)$ is an approximation to $h_0$ obtained by selecting
the $s$ elements of $h_0$ as those that are larger than $\eta h_{max}$ and setting the remaining elements to
zero. Now we state our main result:

\section{Algorithm Efficiency}
\textit{If the measurement matrix $\Phi\in\mathbb{C}^{m\times d^4}$ is drawn
randomly from a probability distribution that satisfies the concentration
inequality in (\ref{RIP}) with $\delta_s<\sqrt{2}-1$, then there exist constants $c_2,c_3,d_1,d_2>0$
such that the solution $h^\star$ to the convex optimization problem,
\begin{align}
&\text{minimize $||h||_{l_1}$}  \notag \\
&\text{subject to $||p^{\prime }-\Phi h||_{l_2}\leq \epsilon$},
\label{decoder}
\end{align}
satisfies,
\begin{align}
||h^\star-h_0||_{l_2} \leq \frac{d_1}{\sqrt{s}} ||h_0(s)-h_0||_{l_1}+d_2%
\epsilon  \label{nearlysparse}
\end{align}
with probability $\geq 1-2e^{-mc_3}$ provided that,
\begin{eqnarray}
m \geq c_2s\log(d^4/s),\label{m}
\end{eqnarray}}where the performance of a $l_1$ minimizer, Eq. (\ref{nearlysparse}), and
the necessary bound $\delta_s<\sqrt{2}-1$ are derived by Cand\'es in Ref. \cite{candes-rip}.

As an example, for a system consisting of $n$ interacting qubits, %M: two-level
%systems, e.g. an $n$ qubit quantum computer,
the exponential number of
parameters describing the dynamics, $2^{2n}$, can be estimated with a
linearly growing number of experiments $m \geq c_2s(8\log(2)n-\log(s))$.
The second term, $d_2\epsilon$, indicates that the algorithmic
performance is bounded by the experimental uncertainties.
Consequently, for fully sparse Hamiltonians and $\epsilon=0$ the
exact identification of an unknown Hamiltonian is achievable.
The properties of the ensemble from which the states and measurement observables
are chosen would determine the parameter $\delta_s$ and consequently the performance
of the algorithm.
The linear independency of the $\Phi$ matrix rows for a random set of local state preparations
 and observables can be guaranteed by a polynomial level of computational overhead before conducting the experiments.

 A certification for the nearly sparsity assumption can be obtained from
 Eqs.(\ref{nearlysparse}) and (\ref{m}) as follows: Suppose $h^\star_m$ is the algorithm's outcome
 for $m$ configurations. The nearly sparsity assumption is certified on the fly during the experiment,
 if the estimation improvement $||h^\star_{m+1}-h^\star_m||$
 converges to zero for a polynomially large total number of configurations.

\section{Physically nearly sparse hamiltonian}

Although physical systems at the fundamental level involve local
two-body interactions, many-body Hamiltonians often describe quantum
dynamics in a particular representation
%(e.g., via a Jordan-Wigner transformation)
 or in well defined approximate limits.
The strength of the non-local $k$-body terms typically is much smaller
than the two-body terms with strength $J$ and decreases with the number $k$.
For a fixed sparsity threshold $\eta$, $k_{\eta}$ is defined as the largest number $k$
for which $k$-body terms have strength larger than $\eta J$.
Then the number of the elements of a $s$-sparse approximation of a $n$-body Hamiltonian
grows linearly as $\mathcal{O}(n g(k_\eta))$, where the $g(k_\eta)$ is determined by the geometry of the system.

A general class of many-body interactions arises when we
change the basis for a bosonic or ferminoic system expressed by a
(typically local) second-quantized Hamiltonian to a Pauli basis,
e.g., via a Jordan-Wigner transformation. For fermionic systems the
interactions are imposed physically from Coulomb's force and Pauli
exclusion principle.
%m
The second-quantized Hamiltonian for these systems can be generally
written as:
\begin{equation}
\hat{H}=\sum_{p,q}b_{pq}\hat{a}_{p}^{+}\hat{a}_{q}+\sum_{p,q,r,s}b_{pqrs}%
\hat{a}_{p}^{+}\hat{a}_{q}^{+}\hat{a}_{r}\hat{a}_{s},
\label{eq:ham2nd}
\end{equation}%
where the annihilation and creation operators ($\hat{a}_{j}$ and $\hat{a}%
_{j}^{+}$ respectively) satisfy the fermionic anti-commutation relations: $\{%
\hat{a}_{i},\hat{a}_{j}^{+}\}=\delta _{ij}$ and $\{\hat{a}_{i},\hat{a}%
_{j}\}=0$ \cite{Mahan}.
%m: and the indices $p$, $q$, $r$, and $s$ run over a given basis.
For example, in chemical systems the coefficients $h_{pq}$ and
$h_{pqrs}$ can be evaluated using the Hartree-Fock procedure for $N$
single-electron basis functions. The Jordan-Wigner transformation
can then be used to map the fermionic creation and annihilation
operators into a representation in terms of Pauli matrices
$\hat{\sigma}^{x},$~$\hat{\sigma}^{y},\hat{\sigma}^{z} $. This
allows for a convenient implementation on a quantum computer, as was
demonstrated recently for the efficient simulation of chemical
energy of molecular systems \cite{Lanyon}. An important example of a
Coulomb based Hamiltonian is the spin-coupled interactions in
quantum dots which has the following Pauli representation:
\begin{equation}
H=\sum_{i,j,k,\cdots }b_{i,j,k,\cdots }\sigma_A^{i}\otimes
\sigma_B^{j}\otimes \sigma_C^ {k}\cdots ,
\end{equation}%
where $A,B,C,\cdots $ indicate the location of the quantum dots,
, $\sigma^i$s are Pauli operators, and $%
b_{i,j,k,\cdots }$ generally represents a many-body spin interacting
term. In practice, these Hamiltonians are highly sparse or almost
sparse %(as defined above)
due to symmetry considerations associated
with total angular momentum \cite{mizel}.
For example the Hamiltonian for the case of four
quantum dots ($A,B,C,D$) takes the general form \cite{mizel}:
\begin{align}
H_{exchange}& =J\sum_{A\leq i<j\leq D}\sigma_{i}.\sigma_{j}+J^{\prime
}[(\sigma_{A}.\sigma_{B})(\sigma_{C}.\sigma_{D})  \notag \\
&
+(\sigma_{A}.\sigma_{C})(\sigma_{B}.\sigma_{D})+(\sigma_{A}.\sigma_{D})(\sigma_{B}.\sigma_{C})],\label{four-body}
\end{align}

Another class of effective many-body interactions often emerge in a perturbative
and/or short time expansion of dynamics, such as effective
three-body interactions between atoms in optical lattices
\cite{Pachos:04} that we study in this work.

Next, we
%M:numerically
simulate the performance of our algorithm for estimation of such
sparse many-body Hamiltonians in optical lattices \cite{Pachos:04}
and quantum dots \cite{mizel}.

\subsection{Three-body interactions in optical lattices}
An optical lattice is a periodic potential formed from interference
of counterpropagating laser beams where neutral atoms are typically
cooled and trapped one per site.
Consider four sites in two adjacent
building blocks of a triangular optical lattice filled by two
species of atoms \cite{Pachos:04}. The interaction between atoms is facilitated by
the tunneling rate $J$ between neighboring sites and collisional
couplings $U$ when two or more atoms occupy the same site. For each site an
effective spin is defined by the presence of one type of atom
as the up-state $\uparrow$ and the presence of the other type as the down-state $%
\downarrow$. Three-body interactions between atoms in a triangular optical
lattice can be significant. The effective Hamiltonian for this
system is studied in Ref. \cite{Pachos:04}. The on-site collisional
interaction $U$, and tunneling rates $J=J^\uparrow=2J^\downarrow$ are
taken to be the same in all sites, also $U=U_{\uparrow\uparrow}=U_{\downarrow%
\downarrow}=2.12U_{\uparrow\downarrow}=10kHz$. The effective Hamiltonian of
the 4-spin system is
\begin{align}
H_{opt-latt}=\sum_{j,\alpha=x,y,z} b_1^\alpha \sigma^\alpha_j\sigma^\alpha_{j+1}
+ b_2^\alpha\sigma^\alpha_j\sigma^\alpha_{j+1}\sigma^\alpha_{j+2}
\end{align}
where $\{b_1^\alpha,b_2^\alpha\}$ are functions of $\{J,U\}$ and their explicit forms are given in appendix D.
The ratio $\eta=|J/U|$ quantifies the sparsity level. For a fixed value of $U$, a smaller $J$ leads to weaker three-body
interactions and therefore a higher level of sparsity. As expected, this
enhances the algorithm performance.

\begin{figure}[htp]
\centering
\includegraphics [width=3.75in] {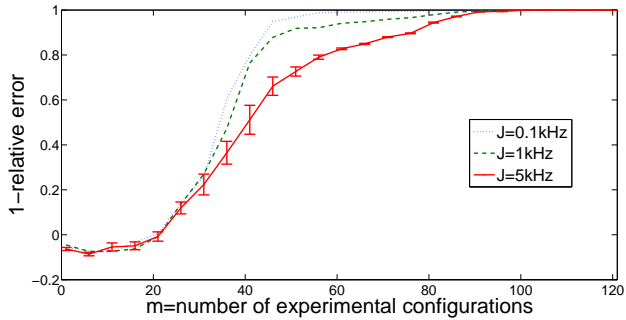}
\caption{The Hamiltonian estimation average performance is illustrated for a system
of four adjacent sites in an optical lattice for different tunneling rates, $J$, and collisional coupling $U=10kHz$.
The error bars demonstrate the standard deviation of the performance
due to the random and independent selection of $m$ configurations (shown only for $J=5kHz$).
Performance accuracy of above $90\%$ with only 60 settings is achievable for $J=1kHz$, which is significantly
smaller than about $6\times10^4$ required experimental configurations in QPT.}
\label{fig_optlattice}
\end{figure}

We assume that the system can be initialized in a random product state $\left|
\psi_{k}\right\rangle=\left| \psi_{k}^1\right\rangle\otimes ...\otimes
\left| \psi_{k}^4\right\rangle$, where $\left| \psi_{k}^i\right\rangle$ are drawn from
the Fubini-Study metric induced distribution. The required observables for the algorithm
are uniformly selected from single qubit Pauli operators $\{\sigma^x_i,\sigma^y_i,\sigma^z_i\}$. This choice
of states and observables allows for $\delta_s\approx 0.37<\sqrt{2}-1$. Let us denote
the extracted Hamiltonian and the true Hamiltonian by $H^*$ and $H_{true}$,
respectively. Here, the performance of the algorithm is defined by the relative error $%
1-||H^*-H_{true}||_{fro}/||H_{true}||_{fro}$. The results for different
number of configurations are depicted in Fig. (\ref%
{fig_optlattice}), for various values of $J$.
As evident in Fig.(\ref{fig_optlattice}), performance accuracy of above
$94\%$ can be obtained with only 80 settings significantly smaller than approximately $6\times10^4$ configurations required in QPT.

%The probabilistic nature of
%the algorithm can be observed from the rough variation of each line.
%The promising advantage of utilizing randomized measurements for parameters
%estimation is evident in Fig. (\ref{fig_optlattice}).
The robustness of this scheme was also investigated for 10\%
random error in simulated experimental data leading to about a 5\% reduction in the overall performance.

\subsection{Four-body interactions in quantum dots}
 Another important class of effective many-body Hamiltonians
can be obtained for electrons in quantum dots coupled through an isotropic
(Heisenberg) or anisotropic exchange interaction.
For example the Hamiltonian for the case of four
quantum dots ($A,B,C,D$) takes the general form Eq. (\ref{four-body}).
The first term in the
summation is a two-body Heisenberg exchange interaction and the last
three terms are four-body spin interactions. %and their strength could
%be as much as $16\%$ of the leading two-body terms .
In certain regimes, the ratio $|J^{\prime }/J|$ can reach up to
$16\%$. The amplitude of $\eta=|J^{\prime }/J|$ determines the sparsity level
of the Hamiltonian.

\begin{figure}[htp]
\centering
\includegraphics [width=3.75in] {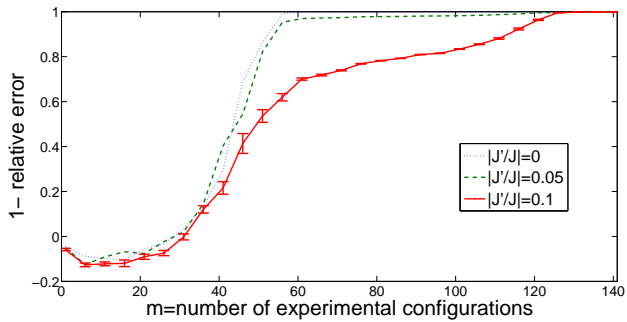}
\caption{Estimation of the exchange interaction Hamiltonian for four
electrons in quantum dots. The average performance of the procedure is illustrated for
different values of $|J^{\prime }/J|$ with $50$ iterations of the $l_1$-reweighted
minimization. The standard deviations are shown only for $|J^{\prime }/J|=0.1$
It is demonstrated that only $60$ different configurations
are sufficient for estimating the unknown Hamiltonian with an
accuracy above $95\%$ for $|J^{\prime }/J|=0.05$, instead of about $6\times 10^4$ required settings via QPT.}
\label{fig_exchange}
\end{figure}

Here we use an efficient
modification of signal recovery referred as "reweighted $l_1$-minimization" which is
described in appendix E. The performance of this algorithm is
demonstrated in Fig. (\ref{fig_exchange}) that shows a significant reduction of the
required number of settings in contrast to the standard QPT.
%Similar to the case of optical lattices above, the minimal number of
%required experimental configurations is about 60 which is significantly
%smaller than $O(6\times 10^4)$ measurement settings prescribed by standard
%QPT.

%In \cite{EPAPS}, we show that for a system interacting
%with a finite-size bath or a surrogate
%bounded Hamiltonian, the system-bath couplings $\{\lambda_{pq}\}$
%can be obtained via a compressed estimation algorithm as:
%\begin{eqnarray}
%p_{jk}&\approx&tr(\rho_k M_{j}) Ê\label{open}\\
%&+&\sum_{pq}\lambda_{pq}tr([\int_0^t ds e^{(t-s)\mathcal{L}_0} \mathcal{L}_{pq} e^{s\mathcal{L}_0}
%[\rho_k],M_j]) \notag
%\end{eqnarray}%
%where $M_j$ is a system only observable and $\mathcal{L}_0$ is
%the system-bath Hamiltonian superoperator. In particular, this approach is
%applicable experimentally when we wish to characterize an effective
%non-unitary quantum control operation on a subsystem of interest
%through a finite-size engineered environment.
%Another method for efficient estimation of magnitude of phenomenological two-body error generators
%in a quantum computer is studied in Ref. \cite{Levi} via monitoring a quantum gate fidelity decay.
%In Ref. \cite{EPAPS}, we also demonstrate how compressed sensing can be applied to Hamiltonian fine structure estimation.
 \section{V. Characterization of Hamiltonian fine structures and system-bath interactions}
\subsection{Hamiltonian fine estimation} In many systems a primary model of
the interactions is often known %a priori
through physical and/or engineering
considerations. Starting with such an initial model we seek to improve
our knowledge about the Hamiltonian by random measurements. Let's assume
the initial guess about the Hamiltonian $H_0$ is close to the true
form $H_{true}$ that is $||\Delta=H_{true}-H_0||\ll ||H_{true}||$.
Therefore for a perturbative treatment we demand $t||\Delta||\ll 1$,
which is a much weaker requirement compared to $t||H_{true}||\ll 1$.
 We can approximate Eq. (1) in the paper to find
\begin{eqnarray}
p_{jk}&\approx&\left\langle \psi_{k}\right|M^0_{j}\left|\psi_{k}\right\rangle
\notag \\
&+&i\left\langle \psi_{k}\right|[\int_0^t e^{isH_0}\Delta e^{-isH_0}
ds,M^0_j]\left|\psi_{k}\right\rangle,  \label{fine}
\end{eqnarray}%
where $M^0_{j}=e^{itH_0} M_j e^{-itH_0}$ \cite{fineref}. This equation is
linear in $\Delta$, consequently, in a similar fashion as above, the
compressed sensing analysis can be applied for efficient estimation of the
fine structure of Hamiltonians.

\subsection{Characterizing system-bath interactions} The identification
of a decoherence process is a vital task for quantum
engineering. In contrast to the usual approach of describing dynamics of
an open quantum system by a Kraus map or a reduce master equation,
here we use a microscopic Hamiltonian picture to efficiently
estimate the system-bath coupling terms generating the overall
decoherence process. However since we consider a full dynamics
of the system and bath, this method can be applied to a finite
size environment such as a spin bath, or a surrogate Hamiltonian
modeling of a infinite bath. In the latter case a harmonic bath of
oscillators is approximated by a finite spin bath \cite{surrogate}.

 Consider an open
quantum system with a total Hamiltonian:
\begin{equation}
H=H_{S}\otimes I_{B}+I_{S}\otimes H_{B}+H_{SB}
\end{equation}
and
\begin{equation}
H_{SB}=\sum_{p,q}\lambda _{p,q}S_{p}\otimes B_{q}
\end{equation}%
where $H_{S}$ ($H_{B}$) denotes the system (bath) free Hamiltonian and $%
H_{SB}$ is the system-bath interaction with coupling strengths $\{\lambda
_{p,q}\}$, and a complete operator basis of the system and bath being $\{S_{p}\}$ and
$\{B_{q}\}$, respectively.

 We develop a formalism to estimate $%
\lambda _{p,q}$ parameters in the weak system-bath coupling regime
and with the sparsity assumption that a few number of $\lambda _{p,q}$ have
a significant value.

The Liouvilian dynamical equation is
\begin{equation}
\frac{d}{dt}\rho_{SB}(t)=(\mathcal{L}_0+\sum_{pq}\lambda_{pq}\mathcal{L}_{pq})[\rho_{SB}(t)]
\label{master}
\end{equation}
where $\mathcal{L}_0[.]=-i[H_{S}\otimes I_{B}+I_{S}\otimes H_{B}, . ]$
and $\mathcal{L}_{pq}[.]=-i[S_{p}\otimes B_{q}, . ]$.
In the regime of weak coupling to a finite bath, $||H_{SB}|| \ll \min\{||H_S||,||H_B||\}$,
the Liouvillan equation (\ref{master}) can be solved perturbatively if time $t$ satisfies $t||H_{SB}|| \ll 1$.
For an initial system density state $\rho_k$, using the matrix identity given in Ref. \cite{fineref} we find the measurement outcomes as
\begin{eqnarray}
p_{jk}&\approx&tr(\rho_k M_{j})  \label{open}\\
&+&\sum_{pq}\lambda_{pq}tr([\int_0^t ds e^{(t-s)\mathcal{L}_0} \mathcal{L}_{pq} e^{s\mathcal{L}_0}
[\rho_k],M_j]) \notag
\end{eqnarray}%
where $M_j$ is a system only observable.
This affine function between the outcomes $p_{jk}$ and coupling parameters $%
\{\lambda_{pq}\}$ is similar to Eq.(2) in the paper for Hamiltonian estimation.
Consequently, the compressed sensing algorithm can be employed for computing
$\{\lambda_{pq}\}$s.

%%%masoud change

\section{Outlook} We have introduced an efficient and
robust experimental procedure for the identification of nearly
sparse Hamiltonians using only separable (local) random state
preparations and measurements. There are a number of future
directions and open problems associated with this work. It is not
known how the performance of the algorithm depends on the
distribution of the ensemble from which the states and measurement
observables are drawn. Also, a general closed-loop learning approach for updating the
knowledge of sparsity basis of an arbitrary Hamiltonian is an interesting open
problem that will be of importance for generic compressed system
identification. The presented method for Hamiltonian estimation is
promising for drastic reduction in the number of experimental
configurations. However the classical resources for post-processing
is not scalable. A fully scalable Hamiltonian estimation method
might be achievable via a hybrid of compressed sensing and DMRG
(Density-Matrix Renormalization Group) methods \cite{Plenio2010}. A
compressed tomography method can also be developed for nearly sparse
quantum processes \cite{Shabani:09}.

%%%end change

\section{Acknowledgement} We thank
NSERC and Center for Extreme
Quantum Information Theory (MM), and
DARPA Grant FA9550-09-1-0710 (RLK, HR) for funding.

\appendix

\section{vectors and operator norm}

In this paper we use the following different norms:

For a vector $x$,
\begin{equation}
||x||_{l_2}=\sqrt{x^\dagger x}, ||x||_{l_1}=\sum_i|x_i|.
\end{equation}

 For a matrix $A$,
 \begin{eqnarray}
% ||A||_{spec}&=&\sqrt{\underset{\left|\psi\right\rangle}{\max}\left\langle\psi\right| A^{\dagger}A\left|\psi\right\rangle/\left\langle\psi|\psi\right\rangle} \\
 ||A||_{spec}&=&\sqrt{\lambda_{max}(A^\dagger A)}
\end{eqnarray}
where $\lambda_{max}$ means largest eigenvalue.
 \begin{eqnarray}
 ||A||_{fro}&=&\sqrt{trace(A^{\dagger}A)}
 \end{eqnarray}

\section{Analysis of the short time approximation}

The short time monitoring of the system's dynamics requires a
prior knowledge of the dynamical time scales.
In the solid-state quantum devices, in particular in the context of
quantum control and quantum information-processing, the time-scale
of single qubit rotations is typically on the order of 1-10 ns. The
switching time for exchange interactions varies among different
solid-state systems. For superconducting phase qubit the duration of
a swap gate is about 10 ns \cite{martinis2010}. For electron-spin
qubits in quantum dots and in donor atoms (Heisenberg models)
\cite{Loss,Loss2,Loss3}, and also for quantum dots in
cavities (anisotropic exchange interactions) \cite{Imamoglu} the
coupling time is between 10-100ps, while for exciton-coupled quantum
dots (XY model) and Forster energy transfer in multichromophoric
complexes the relevant time scale is in the order of 1ps. Next we rigorously
derive bound on the evolution time $t$ that guarantees the
validity of the short time approximation.

For an input state $\left|\psi_k\right\rangle$, the expectation value of an
observable $M_j$ is
\begin{eqnarray}
p_{jk}=\left\langle \psi_{k}(t)\right|M_{j}\left|\psi_{k}(t)\right\rangle
=\left\langle \psi_{k}\right|e^{iHt}M_{j}e^{-iHt}\left|\psi_{k}\right\rangle
\end{eqnarray}

Considering the expansion of the propagator $e^{-iHt}=I-itH-\frac{1}{2}t^2H^2+...$,
we find
\begin{eqnarray}
p_{jk}&=&\left\langle \psi_{k}\right|M_{j}\left|\psi_{k}\right\rangle+it\left\langle \psi_{k}\right|[H,M_{j}]\left|\psi_{k}\right\rangle\notag\\
&-&\frac{t^2}{2}\left\langle \psi_{k}\right|[H,[H,M_j]]\left|\psi_{k}\right\rangle+...
\end{eqnarray}

Therefore, for the linearization assumption, it is sufficient to have for the $l$'th term
\begin{eqnarray}
 t^l \underset{j}{\min}\left\langle \psi_{k}\right|\overbrace{[H,[H,[...}^{l \textrm{ times}},M_j]]]\left|\psi_{k}\right\rangle\leq \notag\\
t^l\underset{j}{\min}|| [H,[H,[...,M_j]]]||_{spec} \ll  1, \forall l.
\end{eqnarray}

A tighter bound can be found for operators $\{M_j\}$ from a POVM as
\begin{equation}
||[H,[H,[...,M_j]]]||_{spec}\leq 2^l ||H||^l_{spec}
\end{equation}
To derive this we use
\begin{equation}
||[A,B]||_{spec}\leq ||AB||_{spec}+||BA||_{spec}\leq 2||A||_{spec}||B||_{spec}
\end{equation}
and $||A||_{spec}^2=||AA^\dagger||_{spec}$.

This gives a single bound sufficient for linearization: $t\ll\frac{1}{2}||H||_{spec}^{-1}$.

\section{RIP from a concentration inequality}
\label{app A}
In this work, we generalize the standard compressed sensing algorithm such that
the necessity for independent randomness in all elements of the measurement matrix, $\phi$, can be avoided.
A common approach to establish RIP (\cite{RIP}) for a matrix $\Phi$
is by introducing randomness in the elements of this matrix. This approach benefits from
measure concentration properties of random matrices.
In classical signal processing each element $\Phi_{jk,\alpha}$ can be
independently selected from a random distribution such as Gaussian or
Bernoulli. Whereas in the Hamiltonian estimation formulation
(Eq. (4) in the paper) there is no freedom for independent selection
of the $\Phi$ matrix elements.

Here we prove the concentration inequality that we
employed for establishing the restricted isometry property.

Though $\Phi$ is a random matrix, because it
is constructed from quantum states and observables of a finite
dimensional system, it is bounded. Thus we are able to apply
\textit{Hoeffding's concentration inequality:}
If $v_{1},...,v_{m}$ are independent bounded random variables such
that $\text{Prob.}\{v_{i} \in [a_{i},b_{i}]\}=1$, then for $S=\sum_{i}v_{i}$,
\begin{eqnarray}
\text{Prob.}\{S-{\bf E}(S)\geq t\}\leq e^{-2t^2/\sum_i (b_i-a_i)^2} \notag\\
\text{Prob.}\{S-{\bf E}(S)\leq -t\}\leq e^{-2t^2/\sum_i (b_i-a_i)^2}\label{app1}
\end{eqnarray}
for any $t>0$. (Here ${\bf E}$ denotes the expectation value.)
Set $v_{i}=|\phi_{i}^\dag x|^{2}$ for a row $\phi_i$. Then with $S=\sum_i v_i=||\Phi x||^2_{l_2}$, we get $\forall x$,
\begin{eqnarray}
v_i=x^\dagger(\phi_i\phi_i^\dagger)x\in(1/m)[w_l,w_u] ||x||^2_{l_2} \notag\\
{\bf E}(S)={\bf E}||\Phi x||^2_{l_2}\in [f,g]||x||^2_{l_2}\label{app2}
\end{eqnarray}
for constants $w_l,w_u,f,g$. Note that $f$ and $g$ are the min
and max singular values of ${\bf E}(\Phi^\dag\Phi)$.
From (\ref{app2}) we find $\forall t_+, t_->0$ and $\forall x$,
\begin{eqnarray*}
\text{Prob.}\{S-g||x||^2_{l_2} \geq t_+\}
&\leq&
\text{Prob.}\{S-{\bf E}(S) \geq t_+\}
\\
\text{Prob.}\{S-f||x||^2_{l_2} \leq -t_-\}
&\leq&
\text{Prob.}\{S-{\bf E}(S) \leq -t_-\}
\end{eqnarray*}
These together with (\ref{app1}) and (\ref{app2}), and the choice of
$t_+=(\delta+1-g)||x||^2_{l_2}$ and
$t_-=(f-1+\delta)||x||^2_{l_2}$ yields
\begin{eqnarray}
\text{Prob.}\{||\Phi x||^2_{l_2}-||x||^2_{l_2}|
\geq \delta ||x||^2_{l_2}\}
\leq 2e^\frac{-2m(\delta+\epsilon)^2}{(w_u-w_l)^2}
\end{eqnarray}
with $\epsilon=\min\{1-g,f-1\}$.  To ensure that $t_+,t_->0$, we
need $1-\delta<f\leq g<1+\delta$. Since the observable $M$ can be
scaled by any real number, a sufficient condition is $g/f <
(1+\delta)/(1-\delta)$. For the simulations in this paper, this ratio becomes $2.176$.

\section{4-sites optical lattice Hamiltonian}
Let us consider four
sites in two adjacent building blocks of a triangular optical
lattice filled by two species of atoms, $\uparrow$ and $\downarrow$.
Atoms interact by tunneling between neighboring sites, $J^\uparrow$
and $J^\downarrow$, and through collisional couplings in the same
site, $U$. The Hamiltonian for such system can be written as
\cite{Pachos:04}:
\begin{align}
H_{opt-latt}=\sum_{j} (0.03\frac{J^{\uparrow 2}+J^{\downarrow 2}}{U}-0.27
\frac{J^{\uparrow 3}+J^{\downarrow 3}}{U^2})\sigma^z_j\sigma^z_{j+1}  \notag \\
-(\frac{2.1(J^\uparrow+J^\downarrow)J^\uparrow J^\downarrow}{U^2}+\frac{%
J^\uparrow J^\downarrow}{U})(\sigma^x_j\sigma^x_{j+1}+\sigma^y_j\sigma^y_{j+1})  \notag \\
+ \sum_{j} 0.14\frac{J^{\uparrow 3}-J^{\downarrow 3}}{U^2}\sigma^z_j\sigma^z_{j+1}\sigma^z_{j+2}
\notag \\
-0.6\frac{J^\uparrow J^\downarrow(J^\uparrow-J^\downarrow)}{U^2}%
(\sigma^x_j\sigma^z_{j+1}\sigma^x_{j+2}+\sigma^y_j\sigma^z_{j+1}\sigma^y_{j+2}),
\end{align}
where $\sigma^{x,y,z}_j$ are Pauli operators.

\section{Reweighted $l_1$-minimization}

In order to simulate our alogrithm performance for estimating the above Hamiltonian we
use an iterative algorithm that outperforms the
standard $l_1$ norm minimization \cite{weighted}. This procedure entails initializing a weight matrix $%
W=I_{d^{2}}$ and a weight factor $\sigma >0$, and repeating the
following steps until convergence is reached:
\begin{align}
&\text{1. Solve for $h$, } \text{minimize $||W h||_{l_1}$}  \notag \\
&\text{subject to $||p^{\prime }-\Phi h||_{l_2}\leq \epsilon$}.  \notag \\
&\text{2. Update weights}  \notag \\
& W=diag(1/(|h_1|+\sigma),...,1/(|h_{d^2}|+\sigma)).
\end{align}
where $h=\textit{vec}(h_i)$ is the Hamiltonian vectorized form.
$\Phi$ is the measurement matrix and $p^\prime$ is the experimental data
with a noise threshold $\epsilon$.


\begin{thebibliography}{99}
\bibitem{RIP2} The common definition of RIP is $(1-\delta_{s/2}) ||h||_{l_2}^{2} \leq ||\Phi h||_{l_2}^{2} \leq
(1+\delta_{s/2})||h||_{l_2}^{2}$ for a $s$-sparse $h$ \cite{CS.intro} which is equivalent to the definition (5) in the paper.
\bibitem{fineref} Equations (\ref{fine}) and (\ref{open}) are direct applications of the
matrix identity $e^{(A+B)t}=e^{At}+\int_0^t e^{A(t-s)}Be^{(A+B)s}ds$.
\end{thebibliography}
\end{document}